\documentclass[12pt]{article}
\usepackage[dvips]{graphicx}
\usepackage{latexsym}
\usepackage{amssymb}

\newcommand{\C}{\mathbb{C}}
\newcommand{\CP}{\mathbb{CP}}

\newcommand{\R}{\mathbb{R}}

\def\ep{{\varepsilon}}

\def\ov{\overline}

\def\p{\partial}

\newcommand{\bea}{\begin{eqnarray}}
\newcommand{\eea}{\end{eqnarray}}

\newcommand{\koniec}{\begin{flushright}  $\Box $ \end{flushright}}
\def\be{\begin{equation}}
\def\ee{\end{equation}}

\def\const{\mbox{const}}

\newcommand{\e}{\textbf{e}}
\def\vv{\varepsilon}

\def\ov{\overline}

\def\p{\partial}

\newcommand{\spp}{\mathbb{S}}
\def\ov{\overline}

\newcommand{\hook}{{\setlength{\unitlength}{11pt}   
                   \begin{picture}(.833,.8)
                   \put(.15,.08){\line(1,0){.35}}
                   \put(.5,.08){\line(0,1){.5}}
                   \end{picture}}}

\newtheorem{theo}{Theorem}[section] 
\newtheorem{prop}[theo]{Proposition}

\begin{document}
\pagestyle{plain}
\title{\vskip -70pt
\begin{flushright}
{\normalsize DAMTP-2010-50}\\
\end{flushright}
\vskip 8pt
{\bf Cosmological Einstein--Maxwell Instantons and Euclidean 
Supersymmetry:\\ Anti--Self--Dual Solutions.}\vskip 3pt}
\author{Maciej\ Dunajski
\thanks{email {\tt m.dunajski@damtp.cam.ac.uk}}\\
{\sl Department of Applied Mathematics and Theoretical Physics} \\
{\sl University of Cambridge} \\
{\sl Wilberforce Road, Cambridge CB3 0WA, UK.} \\
{and}\\
{\sl Institute of Mathematics of the  Polish Accademy of Sciences},\\
{\sl \'Sniadeckich 8, 00-956 Warsaw, Poland}.
\\[8pt]
Jan \ Gutowski \thanks{email {\tt jan.gutowski@kcl.ac.uk}}\\
{\sl 
Department of Mathematics, King' s College London }\\
{\sl Strand, London WC2R 2LS, UK.}\\[8pt]
Wafic \ Sabra \thanks{email {\tt ws00@aub.edu.lb}}
\\[3pt]
{\sl  Centre for Advanced Mathematical Sciences and Physics Department}\\[1pt]
{\sl American University of Beirut, Beirut, Lebanon.}\\[8pt]
Paul \ Tod \thanks{email {\tt paul.tod@sjc.ox.ac.uk}}\\
{\sl The Mathematical Institute, Oxford University  }\\
{\sl 24-29 St Giles, Oxford OX1 3LB, UK.}}
\date{}
\maketitle
\thispagestyle{empty}
\begin{abstract}
We classify super--symmetric solutions of the minimal $N=2$ gauged 
Euclidean supergravity in four dimensions. The solutions
with anti--self--dual Maxwell field give rise to anti--self--dual Einstein 
metrics given in terms of solutions to the 
$SU(\infty)$ Toda equation and more general three--dimensional Einstein--Weyl structures. Euclidean Kastor--Traschen metrics
are also characterised by the existence of a certain super covariantly constant spinor. 
\end{abstract}
\newpage

\section{Introduction} 

The bosonic sector of $N=2$ supergravity (SUGRA) in four dimensions
coincides with the Einstein--Maxwell theory. In \cite{Tod83} all solutions 
which admit a supercovariantly
constant  spinor have been found. 

In this work we  shall classify supersymmetric 
solutions of  Euclidean Einstein--Maxwell equations with 
non--zero cosmological constant. It will be shown that the field 
equations in various branches of our classification reduce to the 
Einstein--Weyl system in three dimensions \cite{Hi,JT85,Dbook}
which is integrable by twistor construction. Some of the Euclidean solutions
arise from analytic continuations of real Lorentzian solutions - for example
the Euclidean analogs of Kastor--Traschen metrics \cite{KS} 
belong to this class - while others do not have Lorentzian counterparts. 
In particular all solutions with anti--self--dual Maxwell field belong to the 
latter class. It turns out (Proposition \ref{prop11})
that the anti--self--duality of the Maxwell
field implies the conformal anti--self--duality (ASD) of the Weyl tensor. In this 
paper we shall focus on constructing all solutions belonging to this ASD class.
The non anti--self--dual solutions will be constructed in \cite{DGST2}.
Some of these have a Lorentzian counterpart \cite{CK, CK1, other_paper, GS, Grov1, 
Grov2}. 

\vskip5pt
In the  second part of this introduction we
shall discuss the Euclidean Einstein--Maxwell theory and explain the 
origin of various sign choices in the Euclidean signature. 
In Section 2  we shall use the two--component spinor calculus to classify all 
supersymmetric solutions. The Killing spinor equations  (\ref{spinor_cond})
contain a continuous 
parameter and we shall show that the Killing spinor gives rise to 
a Killing vector only for one special value of this parameter.
In this symmetric case the metric is given in terms of solutions to the 
$SU(\infty)$ Toda equation (Proposition \ref{prop_toda}). 
For all other values of the parameter the solutions do not in general 
admit an isometry.
They do however admit a conformal retraction (Proposition \ref{propGT1}
and Proposition \ref{prop_interpol}).
In Section 3 we shall characterise the Euclidean Kastor--Traschen solutions
by the existence of a supercovariantly constant spinor with certain additional 
properties (Proposition \ref{propKT}). The solutions constructed in this 
section are not anti--self--dual.

\vskip5pt
There are several  motivations for studying Euclidean 
gauged SUGRA solutions.
From the differential geometric perspective the supersymmetric
solutions constructed in Proposition \ref{propGT1} and Proposition 
\ref{prop_interpol} provide
examples of anti--self--dual Einstein metrics. The point is that
the energy--momentum tensor of the ASD Maxwell field vanishes and the Maxwell
equations decouple from the Einstein equations.
In Euclidean Quantum gravity instantons provide
semi--classical description of black hole creations and in the cosmological 
context  this has been implemented  in \cite{HR, Romans, Ross-Man, MM89}.
Finally solutions of  ungauged ($\Lambda=0$) SUGRA can be used to construct
supersymmetric solutions of Lorentzian minimal SUGRA theories in five
and higher dimensions \cite{DH07, Bobev1, Bobev2}. It remains to be seen
whether solutions to the gauged $D=4$ Euclidean SUGRA admit such lifts.
\newpage
{\bf Acknowledgements.} MD, JG and PT
thank the American University of Beirut
for hospitality when some of this work was carried over.
The work of WS is supported in part by the National Science
Foundation under grant number PHY-0903134. JG is supported by EPSRC grant EP/F069774/1. The authors thank David Calderbank for the interesting comments on the manuscript.
\subsection{Euclidean Einstein--Maxwell equations}
Consider Lorentzian Einstein--Maxwell equations possibly with non-vanishing cosmological constant
\be
\label{lorentzian_EM}
G_{ab}+6\Lambda g_{ab}=-T_{ab}, \qquad d F=0, \qquad d*F=0,
\ee
where
\[
T_{ab}= \frac{1}{2}g_{ab}|F|^2-2F_{ac}{F_b}^c
\]
is the Maxwell energy--momentum tensor\footnote{
Our conventions follow Penrose and Rindler \cite{PR}: $[\nabla_a, \nabla_b]V^d={R_{abc}}^dV^c,
R_{ab}={R_{acb}}^c=6\Lambda g_{ab}-2\Phi_{ab}$, where $\Phi_{ab}$ is the traceless Ricci tensor. Using these conventions in the Riemannian settings implies that the hyperbolic space
$H^4$ has $\Lambda>0$ and the four-sphere $S^4$ has $\Lambda<0$.} and
$|F|^2=F_{cd}F^{cd}$. Swapping the electric and magnetic fields, i. e. 
replacing $F$ by
its Hodge dual $*F$ maps solutions to solutions as the Lorentzian Maxwell E-M tensor is unchanged by this transformation. This can be easily seen in the 
two--component spinor notation \cite{PR} where
\[
T_{ab}=2\phi_{AB}\overline\phi_{A'B'}
\]
and 
the duality transformation is $\ov{\phi}_{A'B'}\rightarrow i\ov{\phi}_{A'B'}$ 
and
$\phi_{AB}\rightarrow -i\phi_{AB}$.
This is no longer the case in the Riemannian signature where
\[
T_{ab}=2\phi_{AB}\tilde\phi_{A'B'}
\]
and the spinors $\phi_{AB}$ and $\tilde{\phi}_{A'B'}$ are independent. The transformation
$F\rightarrow *F$ entails to $\tilde{\phi}_{A'B'}\rightarrow \tilde{\phi}_{A'B'}$
and ${\phi}_{AB}\rightarrow -{\phi}_{AB}$, thus $T_{ab}\rightarrow -T_{ab}$.
This duality transformation can be used to `fix wrong signs' arising from the
analytic continuation of a Lorentzian solution. 

As an example 
consider the Reissner--Nordstr\"om--de Sitter
space--time 
\be
\label{RSDS}
g=-V(r)dt^2+V(r)^{-1}dr^2+r^2(d\theta^2+\sin^2{\theta} d\phi^2),\qquad A=-Q\frac{dt}{r},
\ee
where 
\[
V(r)=1-\frac{2m}{r}+\frac{Q^2}{r^2}-2\Lambda r^2,
\]
$m\geq 0$ and $Q$ are constants
and $F=dA$. Now continue this analytically to the Riemannian signature setting $t=i\tau$
and  assuming that $r$ lies between the middle roots of 
the quartic $r^2V(r)=0$.
We encounter an immediate problem as the potential $A$ is now purely imaginary.
There does not seem to be a universally accepted resolution of this problem, and the way in which one proceeds is dictated by an overall aim of the analytic continuation.
According to Hawking and Ross \cite{HR} one should, at least in the Quantum--Mechanical context, accept that the electrically charged solution has imaginary gauge potential.
Alternatively, especially if our interest lies in classical solutions, one could 
replace $A$ by $iA$ which is real. However this continuation changes the overall
sign of $T_{ab}$ and leads to a `wrong' coupling between the gravitational and electromagnetic fields. The coupling can now be `made right' by replacing
$F\rightarrow *F$, resulting in the Maxwell field $F=-*d(Qr^{-1}d\tau)$.

In this discussion we used `wrong' and `right' in inverted commas, as the energy of the Riemannian Maxwell field is not positive definite, and (unlike in the Lorentzian case) the positivity can not be used to fix the relative sign between $G_{ab}$ and  $T_{ab}$.
The cosmological constant in our example doesn't change under the  
analytic continuation.
In Section \ref{sectionthree} we shall however see a different class of examples 
(Kastor--Traschen cosmological multi black holes \cite{KS}) where the analytic continuation leads to a real solution only if $\Lambda$ changes sign: asymptotically de Sitter Lorentizan
solutions become asymptotically hyperbolic Riemannian solutions. 

To avoid making the various sign choices we shall simply look for real solutions of the Euler--Lagrange equations arising from the Lagrangian density
\[
{\cal L}=\sqrt{g}(R+\gamma |F|^2+\delta),
\]
where $\gamma$ and $\delta$ are real constants. The cosmological constant can then be read--off from $\delta$ and the sign of the Maxwell energy--momentum tensor can be adjusted if necessary replacing $F$ by its Hodge dual as explained above. 
\section{Supersymmetric solutions with ASD Maxwell field}
Let the two--form $F$ be an anti--self--dual (ASD) Maxwell field on a
Riemannian four--manifold $(M, g)$, i. e.
\[
d F=0, \qquad *F=-F.
\]
We shall make use of an isomorphism
\[
\C\otimes T M\cong {\spp}\otimes {\spp'}
\]
where the  complex rank--two vector bundles
$\spp, \spp'$  (called spin-bundles) over $M$
are equipped with parallel symplectic
structures $\ep, \ep'$ such that $g=\vv\otimes \vv'$.
We use the standard convention \cite{PR, Dbook} in which spinor indices are capital
letters, unprimed for sections of $\spp$ and primed for sections of $\spp'$.
The two component spinor formalism will be adapted to the Riemannian signature, where the spinor conjugation
preserves the type of spinors. Thus if $\alpha_{A}=(p, q)$ we can
define $\hat{\alpha}_A=(\overline{q}, -\overline{p})$ so that
${\hat{\hat{\alpha}}}_A=-\alpha_A$. This Hermitian conjugation
induces a positive inner product
\[
\hat{\alpha}_A{\alpha}^A=\ep_{AB}\hat{\alpha}^A{\alpha}^B=|p|^2+|q|^2 \,.
\]
We define the inner product on the primed spinors in the same way.
Here $\vv_{AB}$ and $\vv_{A'B'}$  are covariantly constant symplectic
forms with $\vv_{01}=\vv_{0'1'}=1$. These are used to raise and lower spinor indices
according to $\alpha_B=\ep_{AB}\alpha^A, \alpha^B=\ep^{BA}\alpha_A$, and
similarly for primed spinors. 

The spinor decomposition 
of the Riemann tensor is
\begin{eqnarray} \label{riemann}
R_{abcd} &=& \psi_{ABCD} \ep_{A'B'} \ep_{C'D'} + {\psi}_{A'B'C'D'}
\ep_{AB} \ep_{CD} \nonumber\\ 
&&+ \Phi_{ABC'D'} \ep_{A'B'}\ep_{CD} +
\Phi_{A'B'CD} \ep_{AB}\ep_{C'D'} \nonumber\\&& + 2\Lambda (\ep_{AC}\ep_{BD}
\ep_{A'B'} \ep_{C'D'} - \ep_{AB} \ep_{CD} \ep_{A'D'} \ep_{B'C'}),
\end{eqnarray}
where $\psi_{ABCD}$ and $\psi_{A'B'C'D'}$ are ASD and SD Weyl spinors
which are symmetric in their indices, 
$\Phi_{A'B'CD}=\Phi_{(A'B')(CD)}$ is the traceless Ricci spinor and $\Lambda=R/24$ is the cosmological constant.

Making use of the isomorphism $\Lambda^2_-\cong \spp\odot \spp$
we can write $F_{ab}=\phi_{AB}\varepsilon_{A'B'}$ where the symmetric spinor 
$\phi_{AB}=\hat{\phi}_{AB}$ satisfies the ASD Maxwell equations
\[
{\nabla^A}_{A'} \phi_{AB}=0.
\]

Consider the Killing spinor equations \cite{GHull,Tod83}
\begin{eqnarray}
\label{spinor_cond0}
\nabla_{AA'}\alpha_B+c_0 A_a\alpha_B+(c_1\phi_{AB}+c_2\vv_{AB})\beta_{A'}&=&0,\nonumber\\
\nabla_{AA'}\beta_{B'}+c_3 A_a\beta_{B'}+c_4\vv_{A'B'}\alpha_A&=&0,  
\end{eqnarray}
where $A_a$ is a real one--form and
$c_0, \dots, c_4$ are some constant coefficients which we shall now 
determine.  Differentiating (\ref{spinor_cond0}) covariantly,
commuting covariant derivatives and using the 
spinor Ricci identities 
\be
\label{cc1}
{\nabla^A}_{(A'}\nabla_{B')A}\alpha_B+\Phi_{A'B'AB}\alpha^A=0,
\ee
\be
\label{cc2}
{\nabla^{A'}}_{(A}\nabla_{B)A'}\beta_{B'}+\Phi_{A'B'AB}\beta^{A'}=0,
\ee
\be
\label{cc3}
{\nabla^{A'}}_{(A}\nabla_{B)A'}\alpha_{C}+\psi_{ABCD}\alpha^D- 2\Lambda\alpha_{(A}\vv_{B)C}=0,
\ee
\be
\label{cc4}
{\nabla^{A}}_{(A'}\nabla_{B')A}\beta_{C'}+\psi_{A'B'C'D'}\beta^{D'}- 2\Lambda\beta_{(A'}\vv_{B')C'}=0,
\ee
leads 
to the  compatibility conditions:
Equations ({\ref{cc1}}) give
\[
\Phi_{ABA'B'}=0, \qquad c_0{\nabla^A}_{(A'}A_{B')A}=0, \qquad c_0=c_3.
\]
Equations ({\ref{cc2}}) give $F=2dA$ if
${\nabla^A}_{(A'}A_{B')A}=-\tilde{\phi}_{A'B'}, 
{\nabla^{A'}}_{(A}A_{B)A'}=-{\phi}_{AB}
$ or
\[ 
c_3=-c_1c_4.
\]
Equations ({\ref{cc3}}) give
\[
c_2c_4=-\Lambda.
\]
Finally equations ({\ref{cc4}}) give
\[
\psi_{A'B'C'D'}=0
\]
so we deduce
\begin{prop}
\label{prop11}
A Riemannian four--manifold which admits a solution to 
the Killing spinor equations
{\em(\ref{spinor_cond0})} with anti--self--dual Maxwell field $F$ 
is anti--self--dual and Einstein.
\end{prop} 
\vskip10pt
The case $c_0=c_1=c_3=0$ leads to  
some non--trivial  
solutions in $(2, 2)$ signature, but not in $(4, 0)$ so we shall not 
consider it.
If $c_0\neq 0$, then we can redefine $A_a$, $\phi_{AB}$ and $\beta_{A'}$ by rescalings to get rid
of some constants. Set $c_0=-ce^{i\theta}$, where $c$ and $\theta$ are real. The resulting equations are
\begin{eqnarray}
\label{spinor_cond}
\nabla_{AA'}\alpha_B&=&e^{i\theta}A_a\alpha_B+\Big(\frac{e^{i\theta}}{\Lambda}
\phi_{AB}-\vv_{AB}\Big)\beta_{A'}
\\
\nabla_{AA'}\beta_{B'}&=&e^{i\theta} A_a\beta_{B'}+\Lambda\vv_{A'B'}\alpha_A, \nonumber 
\end{eqnarray}
together with equations for spinor conjugates 
\begin{eqnarray}
\label{spinor_cond1}
\nabla_{AA'}\hat{\alpha}_B&=&e^{-i\theta}A_a\hat{\alpha}_B+\Big(\frac{e^{-i\theta}}{\Lambda}\phi_{AB}-
\vv_{AB}\Big)\hat{\beta}_{A'}
\\
\nabla_{AA'}\hat{\beta}_{B'}&=&e^{-i\theta} A_a\hat{\beta}_{B'}+\Lambda\vv_{A'B'}\hat{\alpha}_A.  \nonumber
\end{eqnarray}
\subsection{$\theta=\pi/2$ and $SU(\infty)$ Toda equation}
Now we shall consider the case $\theta=\pi/2$ and show that the resulting metric is the most general ASD Einstein metric with symmetry, and can be found from solutions to $SU(\infty)$ Toda equation. 
\begin{prop}
\label{prop_toda}
Let the Riemannian four--manifold $(M, g)$ admit a solution to the 
Killing spinor equations {\em(\ref{spinor_cond})} with $\theta=\pi/2$ such that
$F_{ab}=\phi_{AB}\vv_{A'B'}$ is an ASD Maxwell field with $F=2dA$. Then $g$ satisfies
ASD Einstein equations with non--zero $\Lambda$. Moreover $g$ admits a Killing vector
and local coordinates $(x, y, z, \tau)$ can be chosen so that
\be
\label{prop1_metric}
g=\frac{1}{z^2}\Big(V(dz^2+e^u(dx^2+dy^2))+V^{-1}(d\tau+\omega)^2\Big),
\ee
where $u=u(x, y, z)$  is a solution of the $SU(\infty)$ Toda equation
\be
\label{todaeq}
u_{xx}+u_{yy}+({e^u})_{zz}=0,
\ee
the function $V$ is given by $-4\Lambda V=zu_z-2$, and $\omega$ is a 
one--form such that
\be
\label{domega}
d\omega=-V_xdy\wedge dz-V_y dz\wedge dx -(Ve^u)_zdx\wedge dy.
\ee
\end{prop}
We have already shown that $g$ is ASD and Einstein. Once we establish
the existence of a symmetry, we could refer to results of Tod \cite{T97} and Przanowski
\cite{Prz} to deduce the canonical form of the metric (\ref{prop1_metric}). 
In the proof below we shall however give
a direct derivation of this form using the Killing spinor equations.

{\bf Proof.} Define two real non--zero functions $U, \widetilde{U}$ by
\be
\label{UUbar}
U=(\varepsilon_{AB}\hat{\alpha}^A\alpha^B)^{-1}, \qquad \widetilde{U}=(\varepsilon_{A'B'}\hat{\beta}^{A'}\beta^{B'})^{-1}.
\ee
Consider a complex tetrad of one--forms 
\be
\label{tetrad_1}
K_a=i(\hat{\alpha}_A\beta_{A'}+{\alpha}_A\hat{\beta}_{A'}),
\quad X_a=\hat{\alpha}_A\beta_{A'}-{\alpha}_A\hat{\beta}_{A'}, \quad
Z_a=\alpha_A\beta_{A'}.
\ee
The one--forms $X=X_ae^a, K=K_ae^a$ are real and the one--form
$Z=Z_ae^a$ is complex. Using the Killing spinor equations (\ref{spinor_cond}) and their conjugations (\ref{spinor_cond1})
we find
\be
\label{dK}
\nabla_aK_b=\vv_{A'B'}(\Lambda(\hat{\alpha}_A\alpha_B+\alpha_A\hat{\alpha}_B)-\frac{i}{\Lambda\widetilde{U}}\phi_{AB})-\varepsilon_{AB}(\hat{\beta}_{A'}\beta_{B'}+\beta_{A'}\hat{\beta}_{B'})
\ee
so that $\nabla_{(a} K_{b)}=0$ and $K$ is a Killing vector. Moreover we find
\[
dX=0, \qquad Z\wedge dZ=0
\]
and deduce existence of a local coordinate system $(\tau, \zeta, q, \ov{q})$ on $M$ such that
\[
K^a\nabla_a=\sqrt{2}\frac{\p}{\p\tau}, \quad X=\sqrt{2}d\zeta, \quad  Z=\frac{1}{\sqrt{2}}pdq
\]
for some complex--valued function $p=p(\zeta, q, \ov{q})$.
Therefore the one--form dual to the Killing vector is $K=\Omega(d\tau+\omega)$,
where $\Omega$ and $\omega$ are a function and a one--form on the space of orbits
of $K$ in $M$. Using
\be
\label{epsilons}
\varepsilon_{AB}=U(\hat{\alpha}_A\alpha_B-\hat{\alpha}_B\alpha_A),
\quad
\varepsilon_{A'B'}=\widetilde{U}(\hat{\beta}_{A'}\beta_{B'}
-\hat{\beta}_{B'}\beta_{A'})
\ee
we find the metric to be
\[
g=\vv_{AB}\vv_{A'B'}e^{AA'}e^{BB'}=U\widetilde{U}(d\zeta^2+|p|^2dqd\ov{q}+\Omega^2(d\tau+\omega)^2).
\]
Finally using $K_aK^a=2(U\widetilde U)^{-1}$ and setting $q=x+iy$,
$|p|^2=e^{\phi}$ where
$\phi=\phi(x, y, \zeta)$ is a real--valued function yields
\[
g=U\widetilde{U} (e^\phi(dx^2+dy^2)+d\zeta^2)+\frac{1}{U\widetilde{U}}(d\tau+\omega)^2.
\]
Now we need to find equations for $\phi, U, \widetilde{U}$ and $\omega$.
Using the Killing spinor equations (\ref{spinor_cond}) gives
\be
\label{UtildeU}
\nabla_a\Big(\frac{1}{\widetilde{U}}\Big)=\Lambda X_a, \quad
\nabla_a\Big(\frac{1}{U}\Big)=-\frac{1}{\Lambda}{\phi_A}^CK_{CA'}+X_a.
\ee
Therefore $\widetilde{U}=(\sqrt{2}\Lambda\zeta)^{-1}$, where we absorbed the integration constant
into the definition of $\zeta$. Defining a coordinate $z=\zeta^{-1}$ and setting
\[
U=\sqrt{2}\Lambda z V, \quad \phi=u+4\log{\zeta}
\]
where $V=V(x, y, z), u=u(x, y, z)$
yields the final form of the metric (\ref{prop1_metric}).
We now use $(\ref{UtildeU})$ to find
\[
\phi_{AB}=\frac{2\Lambda}{|K|^2}K^{A'}_B\nabla_{AA'}
\Big(\frac{1}{U}-\frac{\Lambda}{\widetilde{U}}\Big),
\]
and substitute this to (\ref{dK}). This yields $-4\Lambda V=zu_z-2$,
where $u$ satisfies 
the SU$(\infty)$ Toda equation (\ref{todaeq}), and (\ref{domega}) holds.
\koniec
Note that the rescaled metric $\hat{g}=z^2g$ is of the form given by the LeBrun ansatz \cite{lebrun}
because $V$ satisfies the linearised  
$SU(\infty)$ Toda equation. Therefore 
$\hat{g}$ is K\"ahler with vanishing Ricci scalar. 
Scalar--flat K\"ahler metrics are also solutions to Einstein--Maxwell equations
in the Riemannian signature \cite{flaherty}, where the self--dual (SD) 
and anti--self--dual (ASD) parts of 
the Maxwell field are given  by the K\"ahler form $\Omega$ and (half of) 
the Ricci form $\rho$ respectively
\[
F=\Omega+\frac{\rho}{2}.
\]
Note that $\Omega\in \Lambda^2_+$ and $\rho\in\Lambda^2_-$.
Thus there exist two conformally related metrics: one non-supersymmetric $\hat{g}$ which solves Euclidean ungauged supergravity equations (Einstein--Maxwell 
with $\Lambda=0$), 
and one supersymmetric $g$ which solves
the gauged supergravity (Einstein--Maxwell 
with $\Lambda\neq0$).
\vskip10pt
\vskip10pt
\subsection{$\theta=0$ and the hyperCR equation}
The ASD Einstein metrics corresponding to $\theta\neq \pi/2$ in (\ref{spinor_cond}) do not in general 
admit a continuous symmetry. In this subsection we shall find a general local form of the metric in the case
when $\theta=0$.
\begin{prop}
\label{propGT1}
Let the Riemannian four--manifold $(M, g)$ admit a solution to the 
Killing spinor equations {\em(\ref{spinor_cond})} with $\theta=0$ such that
$F_{ab}=\phi_{AB}\vv_{A'B'}$ is an ASD Maxwell field with $F=2dA$. Then $g$ satisfies
ASD Einstein equations with $\Lambda>0$. Moreover a local coordinate $\psi$ can be chosen so that
\be
\label{GTtheta0}
g=\frac{\Lambda}{8}\sinh{(2\psi)}^2h+\frac{2}{\Lambda}(d\psi-\coth{(\psi)}\omega)^2, \qquad F=2\;d(\coth{(\psi)}^2\omega)
\ee
where $h=\e_1^2+\e_2^2+\e_3^2$ is a 3-metric, the $\psi$--independent one--forms $(\e_i, \omega)$ satisfy
$\p_\psi\hook \e_i=\p_\psi\hook\omega=0$, 
\begin{eqnarray}
\label{tod_eq}
d\e_1&=&-2\omega\wedge \e_1-\Lambda \e_2\wedge \e_3,\nonumber\\
d\e_2&=&-2\omega\wedge \e_2-\Lambda \e_3\wedge \e_1,\\
d\e_3&=&-2\omega\wedge \e_3-\Lambda \e_1\wedge \e_2\nonumber,
\end{eqnarray}
and 
\be
\label{om_eq1}
d\omega=\Lambda *_h\omega,
\ee
where $*_h$ is the Hodge operator of $h$.
\end{prop}
{\bf Proof.}
The ASD Einstein equations follow from the integrability conditions as we have already explained.

The gauge freedom 
\[
\alpha_A\rightarrow e^f\alpha_A, \quad \beta_{A'}\rightarrow e^f\beta_{A'}, \quad
A\rightarrow A-df, \qquad\mbox{where}\quad f:M\longrightarrow \R
\]
can be used to set $\widetilde{U}=1$. Consider a tetrad (\ref{tetrad_1}), so that
with our gauge choice
\[
g_{ab}=\frac{U}{2}(4Z_{(a}\ov{Z}_{b)}+K_aK_b+X_aX_b)
\]
and $X_aX^a=K_aK^a=2Z_a\overline{Z}^a=2U^{-1}$ and all other inner product vanish.

The condition $d(\widetilde{U}^{-1})=0$ implies
\[
A_a=\frac{\Lambda}{2}X_a.
\]
We also find
\[
X^a\nabla_a (U^{-1})=2U^{-1}(\Lambda U^{-1}-1),
\]
so that if $\tau$ is a local coordinate for which $X^a\nabla_a=\p/\p\tau$ then
\be
\label{formula_for_U}
U=\Lambda(1+e^{2\tau}c^2),
\ee
where $c$ is a local function on $M$ independent on $\tau$. Now we use the Killing spinor equations (\ref{spinor_cond}) and (\ref{epsilons}) to find
\begin{eqnarray}
\label{dzdx}
dZ&=&(-UX+i(U-\Lambda)K)\wedge Z\nonumber\\
dK&=&-UX\wedge K+2i(U-\Lambda) Z\wedge\overline{Z},
\end{eqnarray}
so regarding $X=\p/\p \tau$ as a vector field
\[
{\cal L}_XZ=-2Z, \qquad {\cal L}_X K=-2K
\]
where ${\cal L}_X$ denotes the Lie derivative along the vector field $X=X^a\nabla_a$.
Therefore we can set
\[
Z=e^{-2\tau}\widetilde{Z},\quad K=e^{-2\tau}\widetilde{K}
\]
where $\widetilde{Z}$ and $\widetilde{K}$ are one--forms which Lie--derive along $X$. 
The one form $X_a$ is given by $X=2U^{-1}(d\tau+\Omega)$, where $\Omega$ is a one--form
which in general can depend on $\tau$.
We now have to consider two cases
\begin{enumerate}
\item $U=\Lambda$, which corresponds to  vanishing of the function $c$ in (\ref{formula_for_U}). Now
\[
d\widetilde{Z}=-2\Omega\wedge\widetilde{Z}, \quad d\widetilde{K}=-2\Omega\wedge\widetilde{K},
\]
so that $\Omega$ is independent on $\tau$. Taking the exterior derivatives of these
equations gives the integrability condition $d\Omega=0$. Therefore locally there exist
 $\tau$--independent real valued functions $(\phi, x, y, z)$
such that 
\[
\Omega=d\phi, \quad \widetilde{Z}= \frac{1}{2}e^{-2\phi}(dx+id y), \quad
K=e^{-2\phi} d z.
\]
Finally setting $\tilde{\tau}=\tau+\phi$ gives the 
hyperbolic metric
\be
\label{hyperbolic_g}
g=\frac{\Lambda}{2}e^{-4\tilde{\tau}}(dx^2+dy^2+dz^2)+\frac{2}{\Lambda}d\tilde{\tau}^2
\ee
and the vanishing Maxwell field $F=0$. This metric  has $\Lambda>0$ which is consistent with our curvature conventions.
\item Now assume $U\neq\Lambda$. Equations (\ref{dzdx}) imply
\begin{eqnarray}
\label{neweq}
d\widetilde{Z}&=&(-2\Omega+i\Lambda c^2\widetilde{K})\wedge\widetilde{Z}\nonumber\\
d\widetilde{K}&=&-2\Omega\wedge\widetilde{K}+2i\Lambda c^2\widetilde{Z}\wedge\overline{\widetilde{Z}}.
\end{eqnarray}
We can redefine coordinates to set $c=1$. To see it put
\be
\label{new_eqqq}
\widetilde{Z}=\frac{c^{-2}}{2}(\e_1+i\e_2), \quad \widetilde{K}=c^{-2}\e_3, \quad
\tilde{\tau}=\tau-\log{c}, \quad {\omega}=\Omega+d\log{c}.
\ee
Then the metric is given by
\[
g=\Lambda(1+e^{2\tilde{\tau}})\Big(\frac{1}{2}e^{-4\tilde{\tau}}{h}+\frac{2}{\Lambda^2(1+e^{2\tilde{\tau}})^2}(d\tilde{\tau}+{\omega})^2\Big),
\]
where $h=\e_1^2+\e_2^2+\e_3^2$. 
Substituting (\ref{new_eqqq}) into (\ref{neweq}) gives the system
(\ref{tod_eq}) for the one--forms $\e_i$ .The Maxwell
field is given by
\[
F=d\Big(\frac{2\omega}{1+e^{2\tilde{\tau}}}\Big)
\]
and the anti--self--duality condition $F=-*F$ yields (\ref{om_eq1}).
This is also the integrability condition for (\ref{tod_eq}).
Setting $\psi=-\mbox{arctanh}(\sqrt{1+e^{2\tilde{\tau}}}) $ yields
the form of the metric and the Maxwell field given in the statement of the 
Proposition.
\end{enumerate}
\koniec
{\bf Remarks}
\begin{itemize}
\item
Making an analytic continuation $\psi\rightarrow i\psi$ and
changing the sign of $\Lambda$ leads to an ASD Einstein metric given in terms
of trigonometric (rather than hyperbolic) functions. Setting $\Lambda=-4$ 
yields
\[
g=\frac{1}{4}\sin^2{(2\psi)}h+\frac{1}{4}(d\psi+\cot{\psi}\;\omega)^2
\]
\item A three--manifold admitting a system of one--forms $(\e_i, \omega)$
satisfying equations (\ref{tod_eq}) and (\ref{om_eq1}) admits a hyper--CR Einstein--Weyl
(EW) structure \cite{GT98}. There is a well--known construction \cite{JT85} which associates 
ASD conformal structures with symmetry to any EW structure. Proposition \ref{propGT1}
reveals another connection between the hyperCR EW structures and ASD four--manifolds,
where the Einstein metric in an ASD conformal class does not admit a symmetry.
\item In \cite{DT99} it was shown how to reduce the hyperCR conditions
(\ref{tod_eq}) and (\ref{om_eq1}) 
to a single second--order
integrable PDE (which therefore plays a role analogous to the $SU(\infty)$ Toda equation) for one function of three variables.
\end{itemize}
The metric (\ref{GTtheta0}) degenerates at $\psi=0$ but this degeneracy can be absorbed into a conformal factor as
\[
g=\sinh{(2\psi)}^2 \hat{g},\quad \mbox{where}\quad \hat{g}=\frac{\Lambda}{8}h+\frac{2}{\Lambda}(d\chi+e^{-2\chi}\sinh{(2\chi)}\;\omega)^2 
\]
and $\chi=-\mbox{arctanh}{(e^{2\psi})}$. The conformal structure will therefore be regular if 
the pair $(h, \omega)$ is. An example is provided by the Berger sphere, 
where\footnote{The equations (\ref{tod_eq}) hold, but not in the `obvious' frame 
$\e_1=\sigma_1, \e_2=\sigma_2, \e_3=\Lambda\sigma_3$.
See \cite{Chave} for relevant formulae.} 
\[
h=(\sigma_1)^2+(\sigma_2)^2+ \Lambda^2(\sigma_3)^2, \qquad 
\omega=\frac{1}{2}\Lambda\sqrt{1-\Lambda^2}\sigma_3,
\]
where $0<\Lambda\leq 1$ and $\sigma_i$ are the left--invariant one--forms on 
$S^3$ satisfying
\[
d\sigma_1=\sigma_2\wedge\sigma_3, \quad
d\sigma_2=\sigma_3\wedge\sigma_1, \quad
d\sigma_3=\sigma_1\wedge\sigma_2.
\]
\subsection{General $\theta$ and interpolating Einstein--Weyl equations}
Finally we shall analyse the Killing spinor equations 
(\ref{spinor_cond}) and  (\ref{spinor_cond1}), where the parameter
$\theta$ is allowed to take  arbitrary values. Similarly  to the cases $\theta=\pi/2$ and
$\theta=0$ we shall find that the space--time admits a local fibration over a three--dimensional manifold with an Einstein--Weyl structure. The relevant Einstein--Weyl
structure has arisen in \cite{DT99} as the most general symmetry reductions of ASD Ricci--flat equations by a conformal Killing vector.  It contains both the $SU(\infty)$ and hyperCR equations as special cases. The class of
ASD Einstein metrics characterised in the following Proposition  
does not in general 
admit an isometry (or a conformal isometry) unless $\theta=\pi/2$. Instead
it will be shown to admit an ASD conformal retraction in a sense of \cite{Hi} and 
\cite{Cald} (in \cite{Cald} it is
referred to  a conformal submersion. The metrics
from Proposition \ref{propGT1} 
 belong to the class described in Theorem IX in this reference. The metrics characterised by the proposition below appear to be new).
\begin{prop}
\label{prop_interpol}
Let $(M, g)$ be a Riemannian four--manifold which admits a solution
of the Killing spinor equations  {\em(\ref{spinor_cond})} and  {\em(\ref{spinor_cond1})}
such that the two--form $F=2dA$ is anti--self--dual. Then $g$ is 
anti--self--dual and Einstein with $\Lambda\neq 0$ and locally
is of the form
\be
\label{last_metric}
g=
\frac{2}{\Lambda}\Big(
\frac{e^{2\cos{\theta}\tau}}{1+e^{2\cos{\theta}\tau}}\Big)
\Big(d\tau-\omega+\frac{1}{2}\Lambda \tan{\theta}\; e^{2\cos{\theta}\tau}
\; \e_3\Big)^2+ \frac{\Lambda}{2} e^{4\cos{\theta}\tau}(1+e^{-2\cos{\theta}\tau})h
\ee
where $h=\e_1^2+\e_2^2+\e_3^2$ is a 3-metric, the $\tau$--
independent one--forms $(\e_i, \omega)$ satisfy
$\p_\tau\hook \e_i=\p_\tau\hook\omega=0$, and
\begin{eqnarray}
\label{ewequations}
d\e_3&=&-2\cos{\theta} \omega\wedge \e_3
-\Lambda\cos{\theta}\e_1\wedge\e_2\nonumber\\
d(\e_1+i\e_2)&=&(-2e^{-i\theta}\omega
-ie^{-i\theta}\Lambda\e_3)\wedge(\e_1+i\e_2).
\end{eqnarray}
\end{prop}
{\bf Proof.} To establish this result we shall use the same strategy as in the proof of Proposition \ref{propGT1}. The calculations 
are further complicated by the presence of $\theta$
but the main steps are as before: use the gauge freedom to set 
$\widetilde{U}$ to a constant,
explore the Killing spinor equations to solve for the Maxwell potential $A$ and use the
Frobenius theorem to construct a triad of one--forms out of the 
Killing spinors  defining a conformal structure on a three--manifold.

Using (\ref{spinor_cond}) and  (\ref{spinor_cond1})
we find
\begin{eqnarray*}
\nabla_a U^{-1}&=&2U^{-1}\cos{\theta} A_a+\alpha_A\hat{\beta}_{A'}
-\hat{\alpha}_A\beta_{A'}+\frac{1}{\Lambda}\phi_{AB}(e^{-i\theta}
\alpha^B\hat{\beta}_{A'}-e^{i\theta}\hat{\alpha}^B\beta_{A'}),\\
\nabla_a \widetilde{U}^{-1}&=&2\widetilde{U}^{-1}\cos{\theta}A_a+\Lambda
(\alpha_A\hat{\beta}_{A'}
-\hat{\alpha}_A\beta_{A'}),
\end{eqnarray*}
where $U, \widetilde{U}$ are defined by (\ref{UUbar}).
Use the gauge freedom in scalings of the spinors 
to set $\widetilde{U}$ to a constant. This  gives an expression for $A$. Set
\[
T^a=\frac{1}{\sqrt{2}}(e^{-i\theta} \alpha^{A}\hat{\beta}^{A'}-e^{i\theta}\hat{\alpha}^{A}{\beta}^{A'}),
\]
and define a real one--form  $W$ by 
$\alpha_A\hat{\beta}_{A'}=\sqrt{2}^{-1}e^{i\theta}(T_a+iW_a)$, so that 
\[
g(W, W)=g(T, T)=(U\widetilde{U})^{-1}.
\]
We also define two real one--forms $\e_1, \e_2$ by 
 $Z=f\sqrt{2}^{-1} (\e_1+i\e_2)$, where 
$Z^a=\alpha^{A}\beta^{A'}$ and $f$ is some function. 
Now introduce a local coordinate $\tau$ such that 
$T^a\nabla_a=\p/\p\tau$ and so 
\be
\label{form_of_T}
T=(U\widetilde{U})^{-1}(d\tau+\alpha)
\ee
for some one-form $\alpha$ which in general depends on $\tau$ .
Calculating $T^a\nabla_a U^{-1}$ yields
\[
\frac{\p}{\p\tau} U^{-1}= \sqrt{2}\cos\theta\;U^{-1}
(\widetilde{U}^{-1}-\Lambda U^{-1}).
\]
There are two cases to consider: If $U=\lambda\widetilde{U}=\const$, then $\phi_{AB}=0$, so $dA=0$ and without loss of generality we can
set $A=0$ in some gauge. Using an argument analogous to the one leading to 
(\ref{hyperbolic_g}) we find that $g$ is a hyperbolic metric
with constant scalar curvature (this has $\Lambda>0$ in our conventions). 
Otherwise we have
\[
\tau=\frac{1}{\mu}\;
\Big(\ln\frac{{\widetilde{U}}}{(U-\Lambda\widetilde{U})}\Big)+\ln{(c)},
\quad\mbox{where}\quad \mu=\sqrt{2}\widetilde{U}^{-1}\cos{\theta}, \;\;c=\const.\]
The Killing spinor equations give
\[
dZ= 2e^{i\theta} A\wedge Z-2UZ\wedge \bar{Y}-2\Lambda\widetilde{U} 
Z\wedge Y,
\]
where $Y_a=\alpha_A\hat{\beta}_{A'}$.
Set $Z=f(\tau)Z_0$, where 
$\dot{f}/f=\sqrt{2}\exp{(-i\theta)}\widetilde{U}^{-1}$ so that
\[
dZ_0=(\sqrt{2}e^{-i\theta}\widetilde{U}^{-1}\alpha+
i\Big(\frac{\sqrt{2}\Lambda \widetilde{U}}{\cos{\theta}}- 
\sqrt{2}e^{-i\theta}U\Big) W)\wedge Z_0.
\]
We also find
\[
dT+idW= \sqrt{2}e^{-i\theta}(iU\; T\wedge W+ (U-\Lambda \widetilde{U}) 
|f|^2 Z_0\wedge\bar{Z}_0)
\]
and  
\[
dW= \sqrt{2}\cos{\theta}(U\; T\wedge W-
(U-\Lambda \widetilde{U}) |f|^2 \e_1\wedge \e_2).
\]
Defining a one--form $\e_3$ by $W=g(\tau) \e_3$, where $g=g(0)\exp{(\mu\tau)}$ and substituting this 
in the expression for $dW$ yields
\be
\label{de3}
d\e_3=\sqrt{2} \widetilde{U}^{-1}\cos{\theta}\; \alpha\wedge \e_3
-\sqrt{2}\beta \cos{\theta} \e_1\wedge \e_2,
\ee
where $\beta=\widetilde{U}|f(0)|^2/(c g(0))$ is a constant.
Similarly the expression for $dZ_0$ yields
\be
\label{de12}
d(\e_1+i\e_2) =(\sqrt{2} e^{-i\theta}\widetilde{U}^{-1}\alpha
+i\Big(\frac{2\Lambda\widetilde{U}}{\cos{\theta}}-2e^{-i\theta}U\Big)g\e_3)
\wedge(\e_1+i\e_2).
\ee
We now have to establish the dependence of $\alpha$ on $\tau$. The Killing spinor equations yield
\[
\nabla_{(a}T_{b)}= 2\cos{\theta} A_{(a} T_{b)}+\sqrt{2}^{-1}
(\widetilde{U}^{-1}+\Lambda U^{-1})\cos{\theta} g_{ab}.
\]
Therefore ${\cal L}_T h=\theta  \; h$, where $h_{ab}$ is the part of $g_{ab}$
orthogonal to $T^a$ and the last equality is valid modulo $T$.
Thus  $T_a$ is a conformal retraction. Moreover this retraction is ASD in the sense of \cite{Cald} as $dA$ is ASD. We further find
\[
{\cal L}_T T_a= \sqrt{2} \cos{\theta}(\widetilde{U}^{-1}-\Lambda U^{-1})T_a
-\sqrt{2} U^{-1}\sin{\theta}\; W_a.
\]
Finally using (\ref{form_of_T}) gives
\[
\alpha=-\omega+\Lambda \widetilde{U}^{2} g(0) \tan{\theta}\; e^{\mu\tau}\; 
\e_3,
\]
where $\omega$ is some $\tau$--independent one--form orthogonal to $\p/\p\tau$.
To obtain equations (\ref{ewequations}) in the Proposition we  substitute this expression into (\ref{de3}) and (\ref{de12}), and make the following choices
for the so far unspecified constants
\[
\widetilde{U}=\frac{1}{\sqrt{2}}, \quad g(0)=c\Lambda, \quad 
\beta=\sqrt{2}^{-1}\Lambda
\]
which is consistent if we also chose $(f(0))^2=(g(0))^2$. 
Note that $c$ can also be chosen 
arbitrarily by adding a constant to $\tau$. To obtain the formulae in the Proposition we set $c=\Lambda^{-1}$.
The metric $g$ is given by
\[
g=U\widetilde{U}( T^2+Z^2+|f|^2((\e_1)^2+(\e_2)^2)),
\]
where 
\[
U=\frac{\widetilde{U}+c\widetilde{U}\Lambda e^{\mu\tau}}{c e^{\mu\tau}}.
\]
This, with our choice of constants, gives (\ref{last_metric}).
\koniec
{\bf Remark}. A three--dimensional Einstein--Weyl structure consists
of a conformal structure $[h]$ and a torsion--free connection $D$
such that
\[
D_i h_{jk}=\nu_i h_{jk}, \quad
R_{ij}+\frac{1}{2}\nabla_{(i}\nu_{j)}+\frac{1}{4}\nu_i\nu_j={\cal W} 
h_{jk},\qquad
i, j, k=1, 2, 3,
\]
where $\nu$ is a one--form, 
$\nabla_i$ and $R_{ij}$ are respectively 
the Levi--Civita connection and the Ricci tensor of $h\in [h]$
and ${\cal W}$ is a function which can be read--off by taking a trace of both sides of the second equation. In \cite{DT99} it was shown that
\[
h=(\e_1)^2+(\e_2)^2+ (\e_3)^2, \quad \nu=-4\omega\cos{\theta}
-4\Lambda\sin{\theta} \e_3, \quad \Lambda=\const
\]
satisfies the Einstein--Weyl equations if the triad $(\e_1, \e_2, \e_3)$ 
satisfies (\ref{ewequations}). Moreover the Einstein--Weyl structure arising this way
is the most general symmetry reduction of hyper--K\"aher metric
in four dimension by a conformal symmetry.
\section{Euclidean Kastor--Traschen solutions}
\label{sectionthree}
In this section we shall drop the ASD condition on the Maxwell field so that
\[
F_{ab}=\phi_{AB}\vv_{A'B'}+\tilde{\phi}_{A'B'}\vv_{AB}.
\]
Thus the Killing spinor equations (\ref{spinor_cond0}) are
replaced by 
\begin{eqnarray*}
\nabla_{AA'}\alpha_B+c_0 A_a\alpha_B+(c_1\phi_{AB}+c_2\vv_{AB})\beta_{A'}&=&0,\nonumber\\
\nabla_{AA'}\beta_{B'}+c_3 A_a\beta_{B'}+(c_5\tilde{\phi}_{A'B'}+ c_4\vv_{A'B'})\alpha_A&=&0,  
\end{eqnarray*}
where an additional term involving $\tilde{\phi}_{A'B'}$ is present. We now impose the integrability conditions
(\ref{cc1})--(\ref{cc4}) and proceed as before, but use the Einstein--Maxwell condition
\[\Phi_{ABA'B'}=2\phi_{AB}\tilde{\phi}_{A'B'}.\] We find that
\[
c_0=c_3=\frac{1}{L}, \quad c_2=\frac{1}{2L}c_1, \quad c_5=-\frac{2}{c_1}, \quad c_4=-\frac{1}{Lc_1}, 
\]
where
\[
\Lambda=\frac{1}{2L^2}
\]
is a cosmological constant (which at this stage can be positive or negative if $L$ is real or imaginary respectively).
A constant rescaling of $\beta_{A'}$ can be used to set $c_1$ to any given non--zero constant. 
To achieve a symmetric form of the equations
we replace $\beta_{A'}$ by $\sqrt{2}\beta_{A'}/c_1$  which results in $c_1=\sqrt{2}$. The final form of the
Killing spinor condition is
\begin{eqnarray}
\label{spinor_cond2}
\nabla_{AA'}\alpha_B+\frac{1}{L} A_a\alpha_B+\sqrt{2}(\phi_{AB}+\frac{1}{2L}\vv_{AB})\beta_{A'}&=&0,\nonumber\\
\nabla_{AA'}\beta_{B'}+\frac{1}{L} A_a\beta_{B'} -\sqrt{2}(\tilde{\phi}_{A'B'}+\frac{1}{2L}\vv_{A'B'})\alpha_A&=&0.  
\end{eqnarray}

We conclude that the non-ASD case is more `rigid' than the ASD one. The Killing spinor equations 
(\ref{spinor_cond}) with ASD Maxwell field contain one essential parameter $\theta$. If the Maxwell field is not ASD 
(or SD) the integrability conditions fix all the parameters in terms of the cosmological constant.
Further analysis depends on the sign of the cosmological constant. If $\Lambda<0$, the resulting metric admits a Killing vector
$K_a=i(\hat{\alpha}_A\beta_{A'}+{\alpha}_A\hat{\beta}_{A'})$ and is given by a Riemannian analogue of the Caldarelli--Klemm solution \cite{CK,CK1}.
If $\Lambda>0$ the symmetry is not present in general, and the  metric is a  Riemannian version of the solutions obtained in \cite{GS, other_paper}.

In the proposition below we shall characterise Riemannian Kastor--Traschen solutions \cite{KS} as those where the ratio of norms of the spinors
$\alpha_A$ and $\beta_{A'}$ is a constant. We define two real non--zero functions $U, \widetilde{U}$ by
\[
U=(\varepsilon_{AB}\hat{\alpha}^A\alpha^B)^{-1}, \qquad \widetilde{U}=(\varepsilon_{A'B'}\hat{\beta}^{A'}\beta^{B'})^{-1}.
\]
as before. The gauge transformations $\alpha\rightarrow e^f\alpha, \beta\rightarrow e^f\beta$  where $f:M \longrightarrow \R$
result
in 
\[ 
U\rightarrow e^{-2f} U, \quad \widetilde{U}\rightarrow e^{-2f}\widetilde{U}
\]
so that the ratio $U/\widetilde{U}$ is gauge invariant. In the rest of this section we shall assume $L=l\in\R$ and the cosmological constant is positive.
\begin{prop}
\label{propKT}
Let the Riemannian four--manifold $(M, g)$ admit a solution to the 
Killing spinor equations {\em(\ref{spinor_cond2})} with $\Lambda>0$ such that the gauge invariant condition
\[
\frac{U}{\widetilde{U}}=\const
\]
holds. Then $(M, g)$ is Einstein--Maxwell with\footnote{The sign of the energy momentum tensor in this example is opposite to the one in
(\ref{lorentzian_EM}). This sign can be changed if desired as explained in the Introduction 
by using the Maxwell field $F=2*d(({\cal U}+l^{-1}T)^{-1}dT)$.}
$F=2dA$ and
local coordinates $(x, y, z, T)$ can be chosen so 
that 
\be
g=\Big({\cal U}+\frac{1}{l}T\Big)^2(dx^2+dy^2+dz^2)+\Big(
{\cal U}+\frac{1}{l}T\Big)^{-2}dT^2, \quad
F=2\;d\Big(\frac{dT}{{\cal U}+l^{-1}T}\Big)
\label{KTmetric}
\ee
where ${\cal U}={\cal U}(x, y, z)$ satisfies the Laplace equation on $\R^3$
\[
\frac{\p^2 {\cal U}}{\p x^2}+\frac{\p^2 {\cal U}}{\p y^2}+\frac{\p^2 {\cal U}}{\p z^2}=0.
\]
\end{prop}
{\bf Proof.}
The Einstein--Maxwell equations with $\Lambda>0$ follow from the integrability conditions for (\ref{spinor_cond2}).
To find the local form of the metric first choose a gauge 
\[
U\widetilde{U}=1.
\]
Set $L=l\in\R$. The Killing spinor equations and their conjugates can be used to find
\begin{eqnarray*}
\nabla_a(U^{-1})&=&-\frac{2}{l}A_a U^{-1}-\sqrt{2}\phi_A^BX_{BA'}-\frac{1}{l\sqrt{2}}X_a=0,\\
\nabla_a(\widetilde{U}^{-1})&=&-\frac{2}{l}A_a \widetilde{U}^{-1}-\sqrt{2}{\tilde{\phi}_{A'}}^{B'}X_{AB'}-\frac{1}{l\sqrt{2}}X_a=0.
\end{eqnarray*}
These equations imply
\[
U=\widetilde{U}=1, \quad A_aX^a=-\frac{1}{\sqrt{2}}, \quad 
{{\tilde{\phi}_{A'}}}^{B'}X_{AB'}=\phi_A^BX_{BA'}.
\]
The expression for $A$ is found to be
\be
\label{form_of_A}
A_a=-\frac{l}{2\sqrt{2}}(E_a+\frac{1}{l}X_a),
\ee
where $E_a=2{\phi_A}^BX_{BA'}=X^bF_{ab}$. We also find
\be
\label{EX}
F= E\wedge X.
\ee
Further application of the Killing spinor equations (\ref{spinor_cond2}) gives
\begin{eqnarray}
\label{dzkx}
dZ&=&(-l^{-1}\sqrt{2}X-2l^{-1} A)\wedge Z,\nonumber\\
dK&=&(-l^{-1}\sqrt{2}X-2l^{-1} A)\wedge K,\\
dX&=&-2l^{-1}A\wedge X-\sqrt{2} F,\nonumber
\end{eqnarray}
and consequently
\[
{\cal L}_X Z=-\frac{\sqrt{2}}{l}Z, \quad {\cal L}_X K=-\frac{\sqrt{2}}{l}K, \quad X\wedge dX=0.
\]
The integrability conditions for the first two equations in (\ref{dzkx}) come down to
$d(X+\sqrt{2}A)=0$, so that locally
\be
\label{dxa}
X+\sqrt{2}A=d\gamma
\ee
for some function $\gamma$. Let $\tau$ be a local coordinate such that
$X^a\nabla_a=\p/\p\tau$. We find that $X(\gamma)=1$, so $\gamma=\tau+\tilde{\gamma}$,
where $\tilde{\gamma}$ is a function which does not depend on $\tau$. The
one--form dual to $X=\frac{\p}{\p\tau}$ is $X=2(d\tau+\Omega)$ for some one--form 
$\Omega$. We can fix the falue of the constant $\tilde{\gamma}$ reabsorbing $d\tilde{\gamma}$ into the definition
of $\Omega$. Equations (\ref{dzkx}) now imply the existence of real local coordinates
$(x, y, z)$ such that
\[
K=\frac{1}{\sqrt{2}}e^{-\sqrt{2}\tau/l} dz, \quad Z=\frac{1}{2\sqrt{2}}e^{-\sqrt{2}\tau/l}(dx+idy)
\]
and combining (\ref{form_of_A}) with $X+\sqrt{2}A=d\tau$ yields $\Omega=lE/2$
so that the metric is
\[
g=e^{-2\sqrt{2}\tau/l}(dx^2+dy^2+dz^2)+2\Big(d\tau+l\frac{E}{2}\Big)^2.
\]
Using (\ref{dxa}) and (\ref{EX}) we calculate $dX=l dE=-(\sqrt{2})^{-1}F$ and
\begin{eqnarray*}
{\cal L}_XE&=&-\frac{1}{l\sqrt{2}}X\hook F=-\frac{1}{l\sqrt{2}}X\hook (E\wedge X)\\
&=&\frac{\sqrt{2}}{l} E
\end{eqnarray*}
as $X\cdot X=2$.
Therefore $E=e^{\sqrt{2}\tau/l}\omega$, where $\omega$ is a one--form independent on $\tau$.
The condition $X\wedge dX=0$ implies that $d\omega=0$ so that locally $\omega=d\phi$, where
$\phi=\phi(x, y, z)$ is some function. Using (\ref{dxa}) we find
\[
F=2e^{\sqrt{2}\tau/l}d\phi\wedge d\tau.
\]
Moreover 
\[
*F=\frac{4\sqrt{2}}{l}*_3d\phi,
\]
where $*_3$ is the Hodge operator of the flat 3-metric. Therefore the Maxwell equation
$d*F=0$ implies that $\phi$ is harmonic on $\R^3$ and
\[
g=e^{-2\sqrt{2}\tau/l}(dx^2+dy^2+dz^2)+2\Big(d\tau+\frac{l}{2}e^{\sqrt{2}\tau/l}d\phi\Big)^2.
\]

To put the metric and the Maxwell field
in the form (\ref{KTmetric}) set
\[
T=l\frac{\phi}{\sqrt{2}}-l{e^{-\sqrt{2}\tau/l}}, \quad
{\cal U}=-\frac{\phi}{\sqrt{2}}.
\]
\koniec
The solution (\ref{KTmetric}) can be 
obtained as an analytic continuation of the Kastor--Traschen 
cosmological black holes \cite{KS}. This continuation 
requires the sign of cosmological constant to change. \\ \\
{\bf Example.} Setting ${\cal U}=0$ in (\ref{KTmetric}) gives the hyperbolic space. Consider
${\cal U}=m/R$, where $m$ is a constant, and $R$ is the radial coordinate on $\R^3$ so that 
the metric becomes
\be
\label{rnads}
g=\Big(\frac{m}{R}+\frac{T}{l}\Big)^2\Big(dR^2+R^2(d\theta^2+\sin^2{\theta} d\phi^2)\Big)
+\Big(\frac{m}{R}+\frac{T}{l}\Big)^{-2}dT^2.
\ee
This metric admits an isometry $(R, T)\rightarrow (c^{-1} R, cT)$ generated by the Killing vector 
\[
{\cal K}=T\frac{\p}{\p T}-R\frac{\p}{\p R}.
\]
Introduce the coordinates $(s, r)$ adapted to this isometry by
\[
R=e^{-s/l}, \qquad T=l(r-m)e^{s/l}
\]
so that ${\cal K}(s)=0, {\cal K}(r)=1$. Let $\psi=\psi(s, r)$ be a 
function such that
\[
d\psi=ds+\frac{l(r-m)}{Vr^2}dr,
\]
where 
\[
V=\frac{r^2}{l^2}+\Big(1-\frac{m}{r}\Big)^2.
\]
The metric then takes the form
\be
\label{1instanton}
g=Vd\psi^2+V^{-1}dr^2+r^2(d\theta^2+\sin^2{\theta} d\phi^2).
\ee
It closely resembles the analytic continuation of the Reissner--Nordstr\"om--de Sitter
metric (\ref{RSDS}) described in the introduction in the extremal case
$|Q|=m$. The difference between these two solutions lies in the sign of the 
cosmological constant. The extremal RNdS instanton with $\Lambda<0$
has been named the lukewarm instanton in \cite{Romans}. The conical singularities in the
metric are not present as the black--hole and the cosmological horizons have
the same Hawking temperatures, i.e. $V(r_1)=V(r_2)=0$ at these horizons
and $|V'(r_1)|=|V'(r_2)|$. This instanton has been interpreted
\cite{Romans, Ross-Man} as describing a pair creation of non-extreme black holes
in thermal equilibrium.

In our case $\Lambda>0$. At $r\rightarrow \infty$ the metric approaches the constant curvature hyperbolic space. The limit $r\rightarrow 0$
is singular. 
This reflects the fact that 
the metric (\ref{rnads}) is an analytic continuation of the 
Lorentzian Reissner--Nordstr\"om--de Sitter space--time, where
the singularity is not hidden inside a horizon.
\\ \\
{\bf Example.} If $\Lambda=0$ and 
\[{\cal U}=c+\sum_{m=1}^N\frac{a_m}{|{\bf x}-{\bf x}_m|}, 
\qquad a_1, \dots, a_N, c=\mbox{const}
\]
then (\ref{KTmetric}) becomes the Majumdar--Papapetrou Einstein--Maxwell multi instanton \cite{DH07}. The metric is asymptotically locally Robinson-Bertotti if $c=0$, or asymptotically flat if $c\neq 0$ and $T$ is periodic. 
In \cite{DH07} it was shown how these instantons can be lifted to regular solitonic 
solutions to ${\cal N}=2$ minimal five--dimensional supergravity. It remains to be seen
whether the solutions (\ref{KTmetric}) with non--zero $\Lambda$ can also 
be uplifted to higher dimensions.
\section{Conclusions}
We have classified super--symmetric solutions of the minimal $N=2$ gauged 
Euclidean supergravity in four dimensions, under the additional assumptions that the Maxwell field is anti--self--dual. The resulting metrics
are Einstein, have anti--self--dual Weyl curvature and are given in terms of solutions to three--dimensional 
Einstein--Weyl equations. We have also found one class of examples 
corresponding to non ASD Maxwell field. These examples are  Euclidean 
analogs of Kastor--Traschen cosmological metrics. The solutions 
constructed in the paper 
provide new examples of Einstein metrics in four dimensions. 
It remains to be seen whether they can be used to describe
cosmological black hole creations and in the context of Euclidean Quantum 
Gravity.

\end{document}